# London Penetration Length and String Tension in SU(2) Lattice Gauge Theory

Paolo Cea[a,b,1] and Leonardo Cosmai[b,2]
[a]Dipartimento di Fisica dell'Università di Bari, 70126 Bari, Italy
and
[b]Istituto Nazionale di Fisica Nucleare, Sezione di Bari, 70126 Bari, Italy

**Abstract**

We study the distribution of the color fields due to a static quark-antiquark pair in SU(2) lattice gauge theory. We find evidence of dual Meissner effect. We put out a simple relation between the penetration length and the string tension.

[1]Electronic address: cea@bari.infn.it
[2]Electronic address: cosmai@bari.infn.it

# 1 INTRODUCTION

To shed light on the non perturbative phenomenon of color confinement G. 't Hooft [1] and S. Mandelstam [2] proposed a model known as *dual superconductor model.* According to this model the QCD vacuum behaves as a dual (magnetic) superconductor where the chromoelectric field originated by a $q\overline{q}$ pair is squeezed by Meissner effect into a dual Abrikosov flux tube, giving rise to the confining linear potential.

Up to now there is some some numerical evidence in favour of this model [3, 4, 5, 6]. There are also efforts [7] towards the detection of monopoles condensation in the vacuum, which should cause the formation of the Abrikosov vortices.

# 2 COLOR FIELDS

## 2.1 SU(2)

In order to investigate the field configuration produced by a static $q\overline{q}$ pair we evaluated on the lattice the following connected correlation function [8]

$$\rho_W = \frac{\left\langle \mathrm{tr}\left( W L U_P L^\dagger \right) \right\rangle}{\langle \mathrm{tr}(W) \rangle} - \frac{1}{2} \frac{\langle \mathrm{tr}(U_P) \mathrm{tr}(W) \rangle}{\langle \mathrm{tr}(W) \rangle} , \tag{1}$$

where the plaquette $U_P = U_{\mu\nu}(x)$ in the $(\mu, \nu)$ plane is connected to the Wilson loop $W$ through the Schwinger line L. The relation between $\rho_W$ and the color field stength tensor is [9]

$$F_{\mu\nu}(x) = \frac{\sqrt{\beta}}{2} \rho_W(x) . \tag{2}$$

The color fields distribution can be scanned by varying the position of the plaquette $U_p$ with respect to the Wilson loop $W$.

We performed numerical simulations of the SU(2) lattice gauge theory with Wilson action and periodic boundary conditions. Our data refer to $16^4$, $20^4$ and $24^4$ lattices with $2.45 \leq \beta \leq 2.7$. To reduce the statistical noise we cooled our lattice configurations. We obtained a good signal for $\rho_W$ on very small statistical samples ($20 \div 100$ configurations each one separated by 100 upgrades).

Our results show that $\rho_W$ is sizeable when $U_p$ and $W$ are in parallel planes. This corresponds to measure the component $E_l$ of the chromoelectric field directed along the line joining the $q\overline{q}$ pair. $E_l$ is almost constant along the $q\overline{q}$ line and decrease rapidly in the transverse direction $x_t$. Fig.(1) shows $E_l(x_t)$ measured in the middle of the flux tube for $\beta = 2.7$ on a $24^4$ lattice.

## 2.2 Maximal Abelian Projection

In the 't Hooft's formulation [10] the dual superconductor model is elaborated in the framework of the Abelian projection: after a particular gauge has been fixed the non Abelian gauge theory is described in terms of Abelian projected gauge fields. The physical quantities should be independent of gauge fixing. So we performed measurements of color fields distributions on Abelian projected configurations after maximal Abelian gauge [11] has been fixed. Let $U^A_\mu$ be the Abelian projected links, then the Abelian projected correlator is

$$\rho^A_W = \frac{\left\langle \mathrm{tr}\left(W^A U^A_P\right)\right\rangle}{\langle \mathrm{tr}\,(W^A)\rangle} - \frac{1}{2}\frac{\left\langle \mathrm{tr}\left(U^A_P\right)\mathrm{tr}\left(W^A\right)\right\rangle}{\langle \mathrm{tr}\,(W^A)\rangle}\,. \tag{3}$$

Note that $\rho^A_W$ is a *gauge-dependent* correlator. We performed measurements for $2.45 \leq \beta \leq 2.7$ using $16^4$ and $20^4$ lattices, and we find a good signal without cooling. Measurements are taken on a sample of $500 \div 700$ configurations each separated by 50 upgrades. Fig.(2) shows $E_l(x_t)$ measured in the middle of the flux tube for $\beta = 2.7$ on a $20^4$ lattice. As we can inspect both Fig.(1) and Fig.(2) display the same qualitative behavior.

# 3 LONDON PENETRATION LENGTH

If the dual superconductor scenario holds the transverse shape of $E_l$ is the dual version of the Abrikosov vortex field. Hence we expect that $E_l(x_t)$ can be fitted according to

$$E_l(x_t) = \frac{\Phi}{2\pi}\mu^2 K_0\left(\mu x_t\right)\ ,\quad x_t > 0\,, \tag{4}$$

where $x_t$ is the transverse distance, $K_0$ is the modified Bessel function of order zero, $\Phi$ is the external flux, and $\mu = 1/\lambda$ is the inverse of the London penetration length $\lambda$. Eq.(4) is valid for a type-II superconductor ($\lambda \gg \xi$ : coherence length). We fit Eq.(4) to our data obtaining $\chi^2/f \lesssim 1$ for $x_t \geq 2$ (in lattice units) both in the case of SU(2) gauge configurations and in the case of maximal Abelian projected (m.A.p.) gauge configurations. This is reasonable since Eq.(4) is valid for $\lambda \gg \xi$. We checked the stability of the parameters $\Phi$ and $\mu$ by fitting Eq.(4) to data for $E_l(x_t)$ with $x_t \geq x_t^{min}$ ($x_t^{min}$ =2,3,4,5). Moreover we ascertained that the data obtained from the SU(2) invariant correlator $\rho_W$ on cooled gauge configurations lead to a parameter $\mu$ which shows a plateau vs. the number of cooling steps.

If the London penetration length $\lambda$ is a physical quantity, it should take the same value (within statistical errors) both in the case of SU(2) configurations and in the case of m.A.p. configurations. We have verified that the two values are consistent. Moreover there is an approximate plateau for $\mu/\Lambda_{\overline{MS}}$. Fitting a constant to all data we get $\mu/\Lambda_{\overline{MS}} = 8.8(3)$ with $\chi^2/f = 1.4$.

# 4 STRING TENSION

The string tension is the energy stored into the flux tube per unit length. Since $E_l$ is almost constant along the flux tube we can extimate the string tension $\sigma$ extrapolating Eq.(4) for $x_t = 0$. We get the following relation between the London penetration length $\lambda = 1/\mu$ and the string tension $\sigma$

$$\sqrt{\sigma} = \frac{\Phi}{\sqrt{8\pi}} \mu \ . \tag{5}$$

As explained in Sect. 3 we computed the parameters $\Phi$ and $\mu$ on SU(2) data and on m.A.p. data. In the latter case $\Phi \approx 1$ and independent of $\beta$. On the other hand, for SU(2) $\Phi > 1$ and approaches values close to 1 by increasing $\beta$. We suspect that in this case the external flux $\Phi$ is strongly affected by lattice artefacts which are strongly reduced in the maximal Abelian gauge. We try to get rid of these effects by putting $\Phi = 1$. In this way Eq.(5) becomes

$$\sqrt{\sigma} = \frac{\mu}{\sqrt{8\pi}} \ . \tag{6}$$

In Fig.(3) we can see an approximate plateau for $\sqrt{\sigma}/\Lambda_{\overline{MS}}$. Fitting to a constant we obtain $\sqrt{\sigma}/\Lambda_{\overline{MS}} = 1.76(15)$ which is consistent with the value $\sqrt{\sigma}/\Lambda_{\overline{MS}} = 1.79(12)$ inferred [12] from an asymptotic extrapolation using data for string tension on larger lattices.

In conclusion we have given evidence in favour of dual superconductor model of confinement. We have also shown that that London penetration length $\lambda$, obtained by fitting Eq.(4) to the profile of the chromoelectric flux tube, is a physical quantity related to the string tension.

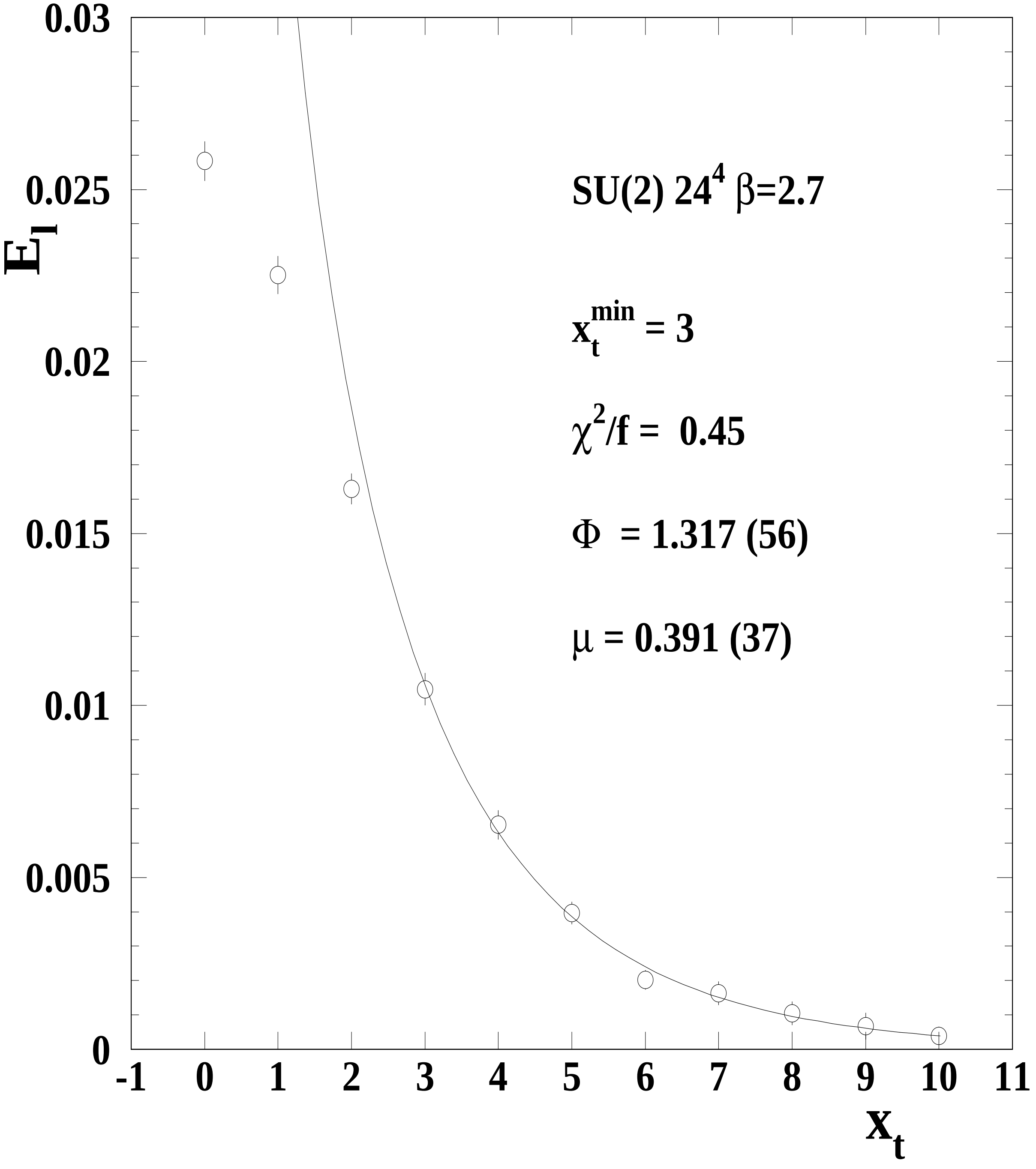

Figure 1: Transverse distribution of the longitudinal color chromoelectric field at $\beta = 2.7$. The size of the Wilson loop $W$ in Eq.(1) is $10 \times 10$. Solid line is the fit Eq.(4).

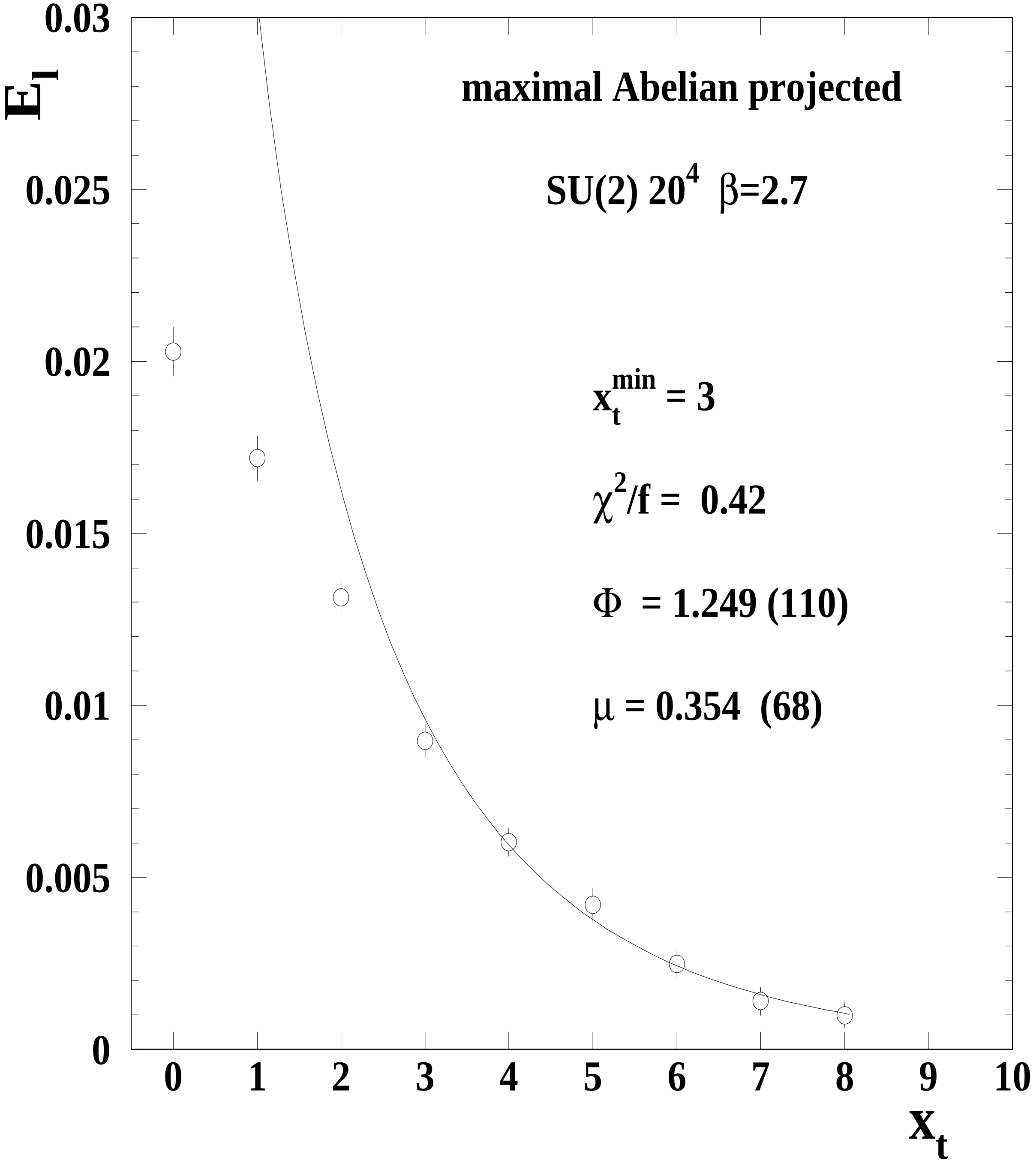

Figure 2: Transverse distribution of the longitudinal color chromoelectric field at $\beta = 2.7$. The size of the Wilson loop $W$ in Eq.(1) is $8 \times 8$. Solid line is the fit Eq.(4).

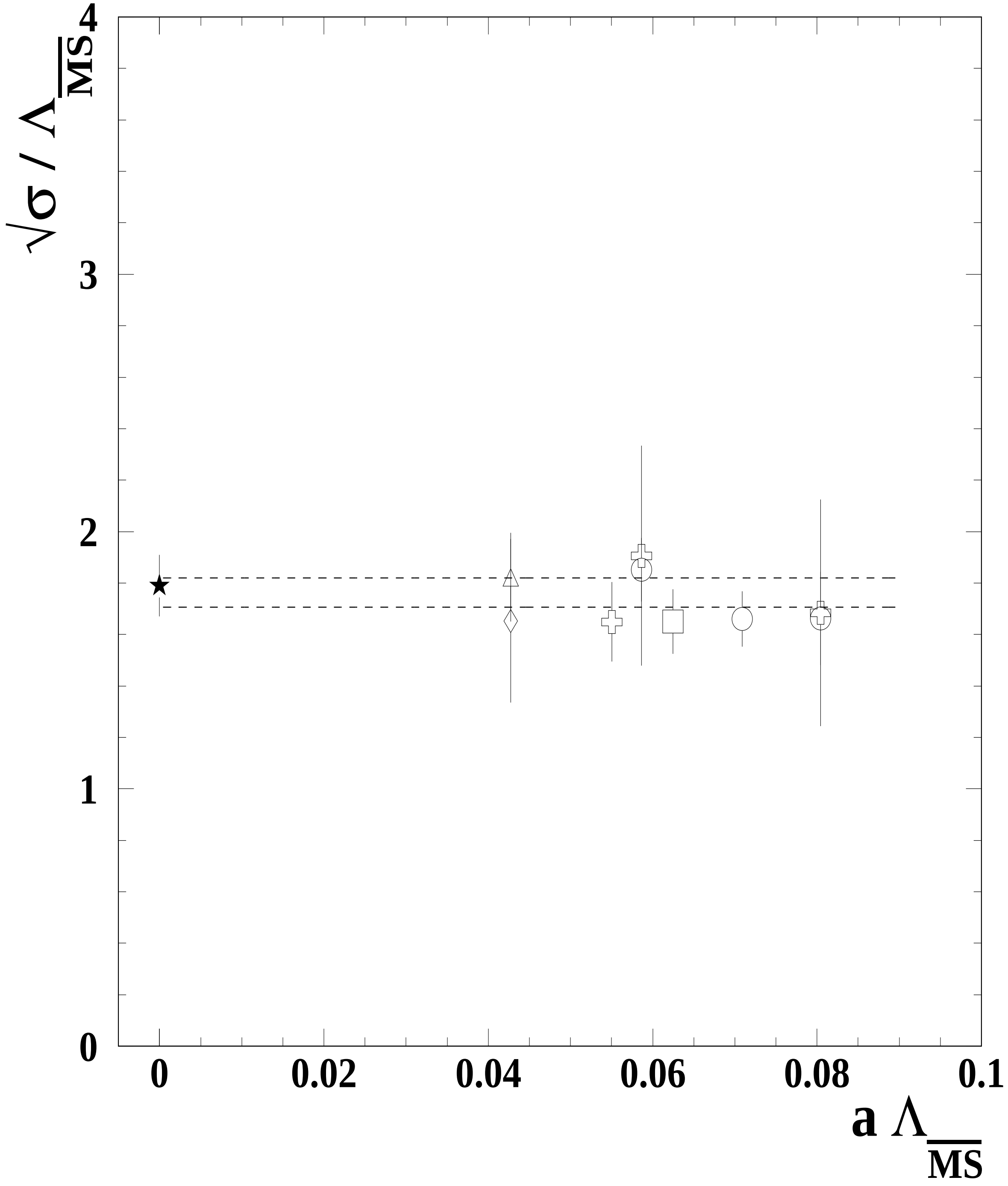

Figure 3: The square root of the string tension in units of $\Lambda_{\overline{MS}}$ versus $a\Lambda_{\overline{MS}}$. Open symbols as in Fig. 3. Full star is the extrapolated continuum limit given in Ref. [12].